\documentclass[10pt]{article}
\usepackage{graphicx}
\usepackage{amsmath}
\usepackage{amssymb}

\begin{document}

\title{SU(5) Symmetry of $spdfg$ Interacting Boson Model}

\author{{LI Jingsheng $^{1)}$, LIU Yuxin$^{1,2,3,4)}$, GAO Peng$^{1)}$} \\
{\small$^1$ Department of Physics, Peking University, Beijing
100871, China}\\
{\small$^2$ The Key Laboratory of Heavy Ion Physics, Ministry
of Education, Peking University, Beijing 100871, China}\\
{\small$^3$ Institute of Theoretical Physics, Academia
Sinica, P.\ O.\ Box 2735,Beijing 100080, China } \\
{\small$^4$ Center of Theoretical Nuclear Physics, National
Laboratory of Heavy Ion Accelerator, Lanzhou 730000, China} }
\date{}
\maketitle

\begin{abstract}
The extended interacting boson model with $s$-, $p$-, $d$-, $f$-
and $g$-bosons being included ($spdfg$ IBM) are investigated. The
algebraic structure including the generators, the Casimir
operators of the groups at the SU(5) dynamical symmetry and the
branching rules of the irreducible representation reductions along
the group chain are obtained. The typical energy spectrum of the
symmetry is given.
\end{abstract}


\hspace*{25mm} Superdeformation, Shape coexistence

\section{Introduction}

\hspace*{6mm}It has been well known that the interacting boson model (IBM)\cite{IA87}, is successful in describing
the spectroscopic properties of low-lying nuclear levels that contain quadrupole collectivity and pairing effects.
In this theory, the pairs of valence nucleons (particles/holes) are approximated as bosons. Then the configuration
space is decreased dramatically with respect to that in shell model. Furthermore, it has been documented that
collective motion is closely related to the dynamical symmetries of the system. For example, in the original
version of IBM, which includes only $s$- and $d$-bosons and those of protons and neutrons are not distinguished
from each other, vibration corresponds to U(5) symmetry, axial rotation to SU(3) symmetry and $\gamma$-unstable
rotation to O(6) symmetry\cite{IA87}. We can then implement algebraic method to study the dynamical properties of
nuclei.

Recently, the nuclear states with octupole deformation which
displays a reflection asymmetric shape have been one of the focus
in nuclear structure studies\cite{BN96,Chen01}. It has been shown
that, with the $p$- and $f$-bosons being included, the interacting
boson model ($spdf$ IBM) can describe the nuclear states with both
quadrupole and octupole deformations
well\cite{EI857,Kus890,Sun91,Liu941,Liu942,JL99}. On the other
hand, to describe the nuclear states with large quadrupole moment
and hexadecupole deformation , the $g$-bosons should be taken into
account\cite{BM802,DK902,L98}. And hence the $sdg$ IBM was put
forward and the algebraic structure has been investigated
well\cite{Sun83,K87}. Meanwhile, it has also been shown that the
$sdg$ IBM is quite powerful to describe superdeformed (SD) states
with positive parity\cite{OH91,Kuy96,Liu97,LSZ97}. The SD states
are those at the extreme condition with large deformation and very
high rotational frequency, and built upon the second well of the
nuclear energy surface. Up to now, several fascinating phenomena,
such as the identical bands, $\Delta I =4$ bifurcation, which
challenge the traditional nuclear many-body theory, have been
observed in SD states. With the IBM being extended to include
$s$-, $p$-, $d$-, $f$-, and $g$-bosons, and to the supersymmetry
formalism, not only the global properties of the positive parity
and negative parity SD bands but also the identical bands, the
$\Delta I =4$ bifurcation can be described well\cite{
Liu97,Liu981,Liu982,Liu99, Liu011,Liu012,Liu013}. Even though the
$spdfg$ IBM is so successful in describing nuclear states, only
the SU(3) symmetry has been investigated thoroughly\cite{Long98}
up to now. Analyzing the algebraic structure of the $spdfg$ IBM in
more detail, one can know that there exists another dynamical
symmetry, namely the SU(5) symmetry, which has been shown to be
more suitable to describe SD states\cite{
Liu97,Liu981,Liu982,Liu99, Liu011,Liu012,Liu013}. Then we will
discuss the SU(5) symmetry of the $spdfg$ IBM in this paper.

\section{Dynamical symmetries of the spdfg IBM}

\hspace*{6mm}In the $spdfg$ IBM, where the $p$-, $f$- and $g$-bosons are included alongside $s$- and $d$-bosons,
the space spanned by the single boson states is $\underset{l}\sum(2 l + 1) = 1 + 3 + 5 + 7 +9 = 25$ dimensional.
The largest symmetry group is then $U(25)$. The generators of the group can be expressed as the Racah tensors
\begin{equation}
G^{k}_{q}(l l^{\prime}) = (b^{\dagger}_{l} \tilde{b}_{l^{\prime}} )^{k}_{q}
= \sum_{\mu \mu ^{\prime}} \langle l \mu l^{\prime} \mu^{\prime} \vert k q
\rangle b^{\dagger}_{l \mu} \tilde{b}_{l^{\prime} \mu^{\prime}} \, ,
\end{equation}
where $b_{l \mu}^{\dagger}$ is the creation operator of the boson with
angular momentum $\{ l, \mu \}$, $\tilde{b}_{l \mu}=(-1)^{l+\mu} b_{l-\mu}$
is the spherical tensor conjugate to $b^{\dagger}_{l\mu}$ with $b_{l\nu}$
being the annihilation operator of the boson with angular momentum $\{ l,
\nu \}$.

To describe the dynamical symmetries of nuclear states consisting
of the $s$-, $p$-, $d$-, $f$- and $g$-bosons, we should consider
group chains starting from U(25) and ending at SO(3) due to the
conservation of angular momentum. It is evident that there exist
maximal subgroups $U_{s}(1) \otimes U_{pdfg}(24)$, $U_{sp}(4)
\otimes U_{dfg}(21)$, $U_{sd}(6) \otimes U_{pfg}(19)$, $U_{sf}(8)
\otimes U_{pdg}(17)$, $U_{sg}(10) \otimes U_{pdf}(15)$,
$U_{spd}(9) \otimes U_{fg}(16)$, $U_{spf}(11) \otimes U_{dg}(14)$,
$U_{sdg}(15) \otimes U_{pf}(10)$, and so on. Because $s$-, $d$-
and $g$-bosons hold positive parity and $p$-, $f$-bosons possess
negative parity, and nuclear states are usually classified with
their parity, we can concentrate our discussion to the largest
subgroup $U_{sdg}(15) \otimes U_{pf}(10)$.

To discuss the dynamical symmetry group chains, one should find
the subgroups of $U_{sdg}(15)$ and $U_{pf}(10)$ at first, then
couple the ones at same level with each other. It has been known
that the group $U_{sdg}(15)$ has subgroups $SU_{sdg}(6)$,
$SU_{sdg}(5)$ and $SU_{sdg}(3)$ \cite{Sun83,K87}, and the group
$U_{pf}(10)$ holds subgroups $SU_{pf}(5)$ and
$SU_{pf}(3)$\cite{Kus890,Sun91,Liu941}. These subgroups can couple
with each other at levels SU(5) and SU(3). Therefore, the $spdfg$
IBM possesses dynamical symmetries SU(5) and SU(3). The SU(3)
symmetry has been investigated in Ref.\cite{Long98}. We discuss
then the SU(5) symmetry in the $spdfg$ IBM in this work.

\section{Algebraic Structure of the $SU_{spdfg}(5)$ Symmetry }

\hspace*{6mm}On the coupling level $SU_{sdg}(5)\otimes SU_{pf}(5) \supset SU_{spdfg}(5)$, we have the dynamical
group chain
\begin{eqnarray}
& U_{spdfg}(25)\supset {U_{sdg}(15)\otimes {U_{pf}(10)}}\supset
SU_{sdg}(5)\otimes SU_{pf}(5) \supset SU_{spdfg}(5) & \nonumber\\&
\supset{SO_{spdfg}(5)}\supset {SO_{spdfg}(3)} \, ,
\end{eqnarray}
The wavefunction can be expressed in terms of the irreducible
representations (irreps) of the groups as:
\begin{eqnarray}
&&\hspace{-0.5cm}\left| {\psi _L^\pi } \right\rangle =\left| {N;\
n_{sd\textsl{g}},\ [n_i^{+}]_{5};\  n_{pf},\ [n_i^{-}]_{5};\
\gamma\ [n_i]_{5};\ \beta \ \ (\nu _1
,\nu _2 )_5;\ \alpha \ \ L} \right\rangle  \nonumber\\
&&\hspace{2cm}SU_{sdg}(5)\ \ SU_{pf}(5)\ \ SU_{spdfg}(5)\ \ \ \
SO_{spdfg}(5)\ \ \ \ \ SO_{spdfg}(3)
\end{eqnarray}
where $[n_i^{+}]_{5}$, $[n_i^{-}]_{5}$ and $[n_i]_{5}$ denote the
irreps $[n_1^{+},n_2^{+},n_3^{+},n_4^{+}]_{5}$,
$[n_1^{-},n_2^{-},n_3^{-},n_4^{-}]_{5}$, \linebreak
$[n_1,n_2,n_3,n_4]_{5}$ of the groups $SU_{sdg}(5)$, $SU_{pf}(5)$
and $SU_{spdfg}(5)$, re4spectively. $(\nu_1 , \nu_2)$ is the irrep
of the group $SO_{spdfg}(5)$. $L$ is the irrep of the
$SO_{spdfg}(3)$, i.e., the angular momentum of the system. $\beta$
is the additional quantum number to distinguish the same
$(\nu_1,\nu_2)_5$ belong to the same $[n_1,n_2,n_3,n_4]_{5}$, and
$\alpha$ is the additional quantum number to distinguish the same
$L$ belong to the same $(\nu_1,\nu_2)_5$. $\gamma$ is the
multiplicity of the irrep $[n_1, n_2, n_3, n_4]$.

The structure of subgroup and the branching rules for each step of
the irrep reductions are the following.

\noindent{\bf (1) $U(25) \supset U_{sd\textsl{g}} (15) \otimes
U_{pf} (10)$ }

With the general principle of the boson realization of unitary
group, we have the generators of the group $U_{sdg}(15)$ as
\begin{equation}
 B(l,{l}')_q ^{k} = [b_l^{\dag}\widetilde{b}_{l}']_q ^{(k )},\ \ \ \ l,{l}' = {0, 2, 4}
\end{equation}
The generators of the group $U_{pf}(10)$ can be given as
\begin{equation}
 F(l,{l}')_q ^{k} = [b_l^{\dag}\widetilde{b}_{l}']_q ^{(k)},\ \ \ \  l,{l}' ={1, 3}
\end{equation}

The branching rule to reduce the irrep $[N]_{25}$ of the $U(25)$
to the irreps of $U_{sdg}(15) \otimes U_{pf}(10)$ is
\begin{equation} [N]_{25} = \sum_{\oplus} [n_{sdg} ]_{15}
\otimes [n_{pf} ]_{10} \, ,
\end{equation}
where $n_{sdg}$ and $n_{pf}$ are the possible non-negative
integers satisfying the relation $n_{sdg} + n_{pf} = N$. It is
obvious that $n_{sdg}$ is the number of bosons with positive
parity and $n_{pf}$ is the negative parity bosons' number, and $N$
is the total number of the bosons.

\noindent{\bf (2) $U_{sdg} (15) \supset SU_{sdg}(5)$ }

It has been known that the system consisting of particles with
angular momentum $\lambda$ holds the symmetry $U(2\lambda+1)$.
Meanwhile, for the system possessing the $U(N)$ symmetry with
$N=\underset{l}\sum{2l+1}$, if a system including $r$ virtual
identical particles with angular momentum $\lambda$ can couple to
angular momentum set $\{l \}$, the group $SU(2\lambda+1)$ is a
subgroup of $U(\underset{l}\sum{2l+1})$, and the generators of the
subgroup $SU(2\lambda+1)$ can be given as\cite{Sun83}
\begin{equation}
\hat{T}_{q}^{k}=\sum\limits_{l{l}^{\prime }}{\sqrt{\frac{2l+1}{2k+1}}\langle {%
\lambda ^{r}l||[a_{\lambda }^{\dag }\widetilde{a}_{\lambda }^{{}}]^{k}||{%
\lambda ^{r}{l}^{\prime }\rangle }}}[b_{l}^{\dag
}\widetilde{b}_{l^{\prime }}^{{}}]_{q}^{k}
\end{equation}

Because the system with two $\lambda=2$ particles can form the
angular momenta $l=0,2,4$, the s-, d-, g-boson system has then the
subgroup $SU_{sdg}(5)$. From Eq.(7), one can get the generators of
this subgroup as
\begin{eqnarray}
\hat{Q}^{(1)}_q(sdg) & = & [d^{\dag}\widetilde{d}]^{(1)} +
\sqrt6[g^{\dag} \widetilde {g}]^{(1)} \, ,
\\
\hat{Q}^{(2)}_q (sdg) & = &
\frac{2\sqrt{5}}{5}[s^{\dag}\tilde{d}+d^{\dag}\tilde{s}]^{(2)}
-\frac{3}{7}[d^{\dag}\widetilde{d}]^{(2)} +\frac{12
\sqrt{5}}{35}[d^{\dag}\tilde{g}+g^{\dag}\tilde{d}]^{(2)}
+\frac{3\sqrt{22}}{7}[g^{\dag}\tilde{g}]^{(2)}\, ,
\\
\hat{Q}^{(3)}_q (sdg) & = &
\frac{8}{7}[d^{\dag}\widetilde{d}]^{(3)} -\frac{3 \sqrt{10}}{7}
[d^{\dag}\widetilde{g}+g^{\dag}\widetilde{d} ]^{(3)}
-\frac{3}{7}\sqrt{11} [g^{\dag}\widetilde{g}]^{(3)} \, ,
\\
\hat{Q}^{(4)}_q (sdg) & = & \frac{2\sqrt{5}}{5}
[s^{\dag}\widetilde{g}+ g^{\dag}\widetilde{s}]^{(4)}
+\frac{4}{7}[d^{\dag}\widetilde{d}]^{(4)}
+\frac{\sqrt{110}}{7}[d^{\dag}\tilde{g}+ g^{\dag}\tilde {d}]^{(4)}
+\frac{\sqrt{715}}{35}[g^{\dag}\tilde{g}]^{(4)} \, ,
\end{eqnarray}
where $\alpha=\pm 1$, $\beta=\pm 1$, the construct constants of
Lie group are independent of the sign of $\alpha $ or $\beta $.

The branching rule for the reduction $U_{sdg}(15) \supset
U_{sdg}(5)$ is\cite{Sun83}
\begin{equation}
[n_{sd\textsl{g}}]_{15}=\sum\limits_{q,r,s,t}{[2n_{sd\textsl{g}}-4q-6r-8s-10t,\
2r+2s+2t,\ 2s+2t,\ 2t]}
\end{equation}
where $s,t,q,r$ are integers satisfying
\[
2n_{sd\textsl{g}}-4q-6r-8s-10t\geqslant 2r+2s+2t\geqslant 2s+2t
\geqslant 2t\geqslant 0
\]

\noindent{\bf (3) $U_{pf}(10)\supset SU_{pf}(5)$ }

Along the same way to handle the $SU_{sdg}(5)$ group, we get the
generators of group $SU_{pf}(5)$ as
\begin{eqnarray}
\hat{Q}_q^{(1)}(pf) & = & \frac{\sqrt{5}}{5} [p^{\dag}
\widetilde{p}]_q^{(1)} + \frac{\sqrt{70}}{5} [f^{\dag}
\widetilde{f}]_q^{(1)} \, ,
\\
\hat{Q}_q^{(2)}(pf) & = &
-\frac{\sqrt{21}}{5}[p^{\dag}\widetilde{p}]_q^{(2)}
+\frac{2\sqrt{6}}{5}[p^{\dag}\widetilde{f} +
f^{\dag}\widetilde{p}]_q^{(2)}
+\frac{\sqrt{6}}{5}[f^{\dag}\widetilde{f}]_\mu^{(2)} \, ,
\\
\hat{Q}_q^{(3)}(pf) & = &
\frac{\sqrt{30}}{5}[p^{\dag}\widetilde{f} +
f^{\dag}\widetilde{p}]_q^{(3)} - \frac{\sqrt{15}}{5} [f^{\dag}
\widetilde {f}]_q ^{(3)} \, ,
\\
\hat{Q}_q^{(4)}(pf) & = &
-\frac{\sqrt{10}}{5}[p^{\dag}\widetilde{f} +
f^{\dag}\widetilde{p}]_q^{(4)} - \frac{\sqrt{55}}{5}[f^{\dag}
\widetilde {f}]_q ^{(4)}\, .
\end{eqnarray}

The branching rule of the reduction $U_{pf}(10)\supset SU_{pf}(5)$
can be given as\cite{Kus890,Sun91,Liu941}
\begin{equation}
[n_{pf}]_{10}=\left\{{\begin{array}{l}
[n_{pf},n_{pf},0,0]\oplus[n_{pf}\textrm{-1},n_{pf}\textrm{-1,1,1}]\oplus[n_{pf}\textrm{-2,}n_{pf}\textrm{-2,2,2}]
\oplus ...
\\ \
....\oplus[\frac{1}{2}n_{pf},\frac{1}{2}n_{pf},\frac{1}{2}n_{pf},\frac{1}{2}n_{pf}]
\ \ \ \ \ \ \ \ \ \ \ \ \ \ \ \ \ \ \ \ \ \ \ \ \ \ \ \ \ \ \ \
n_{pf} \textrm{ is even}
\\ \\ \
[n_{pf},n_{pf},0,0]\oplus[n_{pf}\textrm{-1},n_{pf}\textrm{-1,1,1}]\oplus[n_{pf}\textrm{-2},n_{pf}\textrm{-2,2,2}]\oplus...
\\ \
....\oplus[\frac{1}{2}(n_{pf}\textrm{+1}),\frac{1}{2}(n_{pf}\textrm{+1}),\frac{1}{2}(n_{pf}\textrm{-1}),\frac{1}{2}(n_{pf}\textrm{-1})]
 \ \ \ \ \ \ \ \ \ \ n_{pf} \textrm{ is odd} \\
 \end{array}} \right.
\end{equation}

\hspace*{6mm}

\noindent{\bf (4) $SU_{sdg}(5)\otimes SU_{pf}(5)\supset
SU_{spdfg}(5)$ }

It is obvious that the generators of $SU_{spdfg}(5)$ are
\begin{equation}
\hat{T}_{q}^{(k)}(spdfg) = \hat{Q}_{q}^{(k)}(sdg) +
\hat{Q}^{(k)}_{q}(pf) \, ,
\end{equation}
where $\hat{Q}_{q}^{(k)}(sdg)$ and $\hat{Q}_{q}^{(k)}(pf)$ are
generators of the groups $SU_{sdg}(5)$, $SU_{pf}(5)$,
respectively.

The reduction $SU_{sdg}(5)\otimes SU_{pf}(5)\supset SU_{spdfg}(5)$
obeys \textit{Littlewood} rule and can be given with the Young
tableaux technique\cite{HS87}.

\hspace*{6mm}

\noindent{\bf (5) $SU_{spdfg}(5) \supset SO_{spdfg}(5)$ }

The generators of the group $SO_{spdfg}(5)$ contain the tensors of
rank 1 and 3, they are identified as the odd rank tensors of the
$SU_{spdfg}(5)$

\begin{eqnarray}
\hat{T}^{(1)} & = & \sqrt{\frac{1}{5}} [p^{\dag} \widetilde
{p}]^{(1)} + [d^{\dag} \widetilde {d}]^{(1)} + \sqrt
{\frac{14}{5}} [f^{\dag}\widetilde {f}]^{(1)} + \sqrt {6}
[g^{\dag} \widetilde {g}]^{(1)}
\\
\hat{T}^{(3)} &=& [d^{\dag} \widetilde {d}]^{(3)} -
\frac{3}{8}\sqrt {11} [g^{\dag} \widetilde {g}]^{(3)} - \alpha
\frac{3}{4}\sqrt {\frac{5}{2}} [d^{\dag} \widetilde {g} + g^{\dag}
\widetilde {d}]^{(3)} - \sqrt {\frac{3}{5}} [f^{\dag} \widetilde
{f}]^{(3)}
\nonumber\\
& &- \sqrt {\frac{6}{5}} [p^{\dag} \widetilde {f} + f^{\dag}
\widetilde {p}]^{(3)}
\end{eqnarray}

Using the Young tableaux technique\cite{HS87,Ma98} or the Schur
function method\cite{Wyb74}, the branching rule of this reduction
can be fixed.

\hspace*{6mm}

\noindent{\bf (6) $SO_{spdfg}(5) \supset SO_{spdfg} (3)$ }

The generators of the group $SO_{spdfg}(3)$ are
\begin{equation}
\hat{L}_q^{(1)} = \sqrt{10} \hat{T}^{(1)}_q \, ,
\end{equation}
where $\hat{T}^{(1)}_q$ are the generators of the group
$SO_{spdfg}(5)$.

The irrep reduction of $SO(5) \supset SO(3)$ is not simple.
However, it has been discussed in detail. All the branching rules
can be obtained by referring to computer calculation or
tables\cite{HS87}.

Ending this section, we give all the Casimir operators that
associate with the $SU(5)$ symmetry in table 1.

\begin{table}[htbp]
\caption{Casimir operators of U(25) and its subgroups}
\begin{small}
\begin{center}
\begin{tabular}{|c|c|p{6.2cm}|p{3.3cm}|} \hline
Group & IRRP & \hspace*{25mm} Casimir & \hspace*{6mm} Eigenvalue  \\
\hline $ U_{sdgpf} $ & [$N$]& $C_{1U(25)}=\sum
\limits_{l=0}^4{b_l^+b_{l}'}=\hat{N}$ & $N$   \\
\hline $U_{sdg} (15)$ & [$n_{sdg} $]  & $C_{1U_{sdg} (15)}\! =\!
[s^{+} \tilde {s}] \!+\! \sqrt 5 [d^{+} \tilde {d}]_0^0 \!+ \!3
[g^{+} \tilde {g}]_0^0 $
\newline
 $C_{2U_{sdg} (15)} = \sum\limits_{l,{l}',k}
{B(l,{l}')^k } \cdot B(l,{l}')^k $& ${n_{sdg}}_{\ } $
\par  
${n_{sdg} (n_{sdg} +14)}^{\ }$   \\ \hline $U_{pf} (10)$ &
[$n_{pf} $]  & $C_{1U_{pf} (10)}\! = \! \sqrt 3 [p^{+} \tilde
{p}]_0^0 \! +\! \sqrt 7 [f^{+} \tilde {f}]_0^0 $ \par $C_{2U_{pf}
(10)} = \sum\limits_{l,{l}',k} {F(l,{l}')^k } \cdot F(l,{l}')^k $&
$n_{sdg} $
\par $n_{sdg} (n_{sdg} $+9)   \\ \hline $SU_{sdg} (5)$ & [$n_1^{+}
,n_2^{+}, n_3^{+}, n_4^{+} $]  & $C_{2SU_{sdg} (5)} =
\sum\limits_{k} {\hat{Q}^{(k)}(sdg)} \cdot \hat{Q}^{(k)}(sdg) $ &
$f(n_1^{+}, n_2^{+}, n_3^{+}, n_4^{+} )^{\ast} $ \\
\hline $SU_{pf} (5)$& [$n_1^{-}, n_2^{-}, n_3^{-}, n_4^{-}$ ]  &
$C_{2SU_{pf} (5)} = \sum\limits_{k} {\hat{Q}^{(k)}(pf)} \cdot
\hat{Q}^{(k)}(pf)$ &
$f(n_1^{-}, n_2^{-}, n_3^{-}, n_4^{-} ) ^{\ast}$ \\
\hline $SU_{spdfg}$  & [$n_1, n_2, n_3, n_4 $]  & $C_{2SU_{spdfg}
(5)}=\sum\limits_k \hat{T}^{(k)} \cdot \hat{T}^{(k)} $ &
$f(n_1, n_2, n_3, n_4 )^{\ast}$ \\
\hline $SO_{spdfg} (5)$ & ($\nu _1 ,\nu _2 )$  & $C_{2SO_{spdfg}
(5)} = 2(\hat{T}^{(1)} \cdot \hat{T}^{(1)} + \hat{T}^{(3)}\cdot
\hat{T}^{(3)})$ & $\nu_1(\nu_1\!+\!3)\!+\! \nu_2(\nu_2\!+\!1)$\\
\hline $SO_{sdgpf} (3)$& $L$  & $C_{2 SO_{spdfg} (3)} =
2(\hat{L} \cdot \hat{L})$ & $L(L+1)$   \\
\hline
\end{tabular}
\end{center}
*\textrm{where $f(x_1 ,x_2 ,x_3 ,x_4 ) = x_1 (x_1 + 4) + x_2 (x_2
+ 2) + x_3^2 + x_4 (x_4 - 2) - \frac{1}{5}(x_1 + x_2 + x_3 + x_4
)^2$}
\end{small}
\end{table}

\section{Energy Spectrum of the SU(5) Symmetry}

\hspace*{6mm}For a nucleus with the SU(5) symmetry in the $spdfg$ IBM, the Hamiltonian can be written as
\begin{eqnarray}
H&=&\epsilon
C_{1U_{sd\textsl{g}}(15)}+\epsilon'C_{1U_{pf}(10)}+AC_{2SU_{sdg}(5)}
+ A'C_{2SU_{pf}(5)}+B C_{2SU_{spdfg}(5)}\nonumber
\\& &+ CC_{2SO_{spdfg}(5)}+DC_{2SO_{spdfg}(3)}
\end{eqnarray}
The energy of the state can be expressed in terms of the irreps of
the groups as
\begin{eqnarray}
E(L^\pi)&\! = \! & E_0 \! + \! \epsilon n_{sdg} + {\epsilon }'n_{pf} \nonumber\\
& &\! + A [n_1^{+}(n_1^{+}\!+\! 4)\! + \! n_2^{+}(n_2^{+}\! + \!
2) \! + \! n_3^{+ 2} \! +  \! n_4^{+}(n_4^{+} \! - \! 2) \! - \!
\frac{1}{5}(n_1^{+} \! + \! n_2^{+}\! + \! n_3^{+} \! + \! n_4^{+})^2] \nonumber\\
& & + {A}' [n_1^{-}(n_1^{-}\! + \! 4)\! + \! n_2^{-}(n_2^{-}\! +
\! 2)\! + \! n_3^{- 2} \! + \! n_4^{-}(n_4^{-}\! - \!2)\! - \!
\frac{1}{5}(n_1^{-} \! + \! n_2^{-} \! + \! n_3^{-} \! + \! n_4^{-})^2]\nonumber\\
& & + B [n_1(n_1\! + \! 4) \! + \! n_2(n_2\! + \! 2) \! + \! n_3^2
\! + \! n_4(n_4\! - \! 2) \! - \! \frac{1}{5}(n_1\! + \! n_2\! +
\! n_3 \! + \! n_4)^2]\nonumber\\
& & + C [\nu_1(\nu_1+3) + \nu_2(\nu_2+1)] + D L(L+1)
\end{eqnarray}

To give an example explicitly, we discuss a SU(5) symmetry nucleus
with six valence nucleons,  i.e., with boson number $N=3$.
According to the branching rules of the irrep reductions along the
group chain, the quantum numbers of all the possible states can be
listed in Table 2.
\begin{table}[htbp]
\textbf{\small{Table 2: Examples of the branching rules of the
irreducible representation reduction in the symmetry. }}
\textbf{\small{Part A. positive parity states}}
\begin{center}
\begin{footnotesize}
\begin{tabular}
{|c|c|c|c|p{135pt}|} \hline
$U_{sdg}(15) \otimes U_{pf}(10)$ & $SU_{sdg}(5) \otimes SU_{pf}(5)$
  & $SU_{spdfg}(5)$ & O(5) & \hspace*{20mm} O(3) \\  \hline
$[n_{sdg}] \otimes [n_{pf}]$ & $[n^+]\otimes[n^-]$ & $[n]$
  & $(\nu_1, \nu_2)$ & \hspace*{22mm} $L$ \\  \hline
\raisebox{-27.00ex}[0cm][0cm] {[3]$ \otimes $[0]} &
 \raisebox{-6.50ex}[0cm][0cm]  {[6]$ \otimes $[0]}&
 \raisebox{-6.50ex}[0cm][0cm] {[6]}&(6, 0)& 0,3,4,$6^2$,7,8,9,10,12   \\
\cline{4-5}    &                 &       &(4, 0)& 2,4,5,6,8          \\
\cline{4-5}    &                 &        &(2, 0)& 2,4               \\
\cline{4-5}    &                 &       &(0, 0)& 0                  \\
\cline{2-5}    & \raisebox{-10.50ex}[0cm][0cm] {[42]$\otimes$[0]}&
\raisebox{-10.50ex}[0cm][0cm]   {[42]} &(4, 2)& 0,1,$2^2, 3^2, 4^3,5^2,6^3,7^2,8^2$,9,10 \\
\cline{4-5}    &                 &       & (4, 0)& 2,4,5,6,8          \\
\cline{4-5}    &                 &       &(3, 1)& 1,2,$3^2, 4, 5^2,6$,7      \\
\cline{4-5}    &                 &       &(2, 2)& 0,2,3,4,6          \\
\cline{4-5}    &                 &       &$(2, 0)^2$ & 2,4                \\
\cline{4-5}    &                 &       &(0, 0)& 0                  \\
\cline{2-5}    & \raisebox{-4.50ex}[0cm][0cm]
{[222]$\otimes$[0]}&
\raisebox{-4.50ex}[0cm][0cm]                {[222]}&(2, 2)& 0,2,3,4,6          \\
\cline{4-5}                   &          &       &(2, 0)& 2,4                \\
\cline{4-5}                   &          &       &(0, 0)& 0                  \\
\hline \raisebox{-38.50ex}[0cm][0cm] {[1]$\otimes$[2]}&
\raisebox{-25.00ex}[0cm][0cm] {[2]$\otimes$[22]}&
\raisebox{-10.50ex}[0cm][0cm]   {[42]} & (4, 2)& 0,1,$2^2,3^2,4^3,5^2,6^3,7^2,8^2$,9,10 \\
\cline{4-5}    &              &       &(4, 0) & 2,4,5,6,8         \\
\cline{4-5}    &              &       &(3, 1) & 1,$2,3^2,4,5^2$,6,7   \\
\cline{4-5}    &              &       &(2, 2) & 0,2,3,4,6         \\
\cline{4-5}    &              &       &$(2, 0)^2$ & 2,4           \\
\cline{4-5}    &              &       &(0, 0) & 0               \\
\cline{3-5}    &              &
\raisebox{-10.50ex}[0cm][0cm] {[321]} & (3, 2) & 1,2,3,4,5,6,7,8   \\
\cline{4-5}    &              &      & (3, 1) & 1,$2,3^2,4,5^2$,6,7  \\
\cline{4-5}    &              &      & (2, 2) & 0,2,3,4,6         \\
\cline{4-5}    &              &      & (2, 1) & 1,2,3,4,5    \\
\cline{4-5}    &              &      & (2, 0) & 2,4               \\
\cline{4-5}    &              &      & (1, 1) & 1,3               \\
\cline{3-5}    &              &
\raisebox{-4.50ex}[0cm][0cm]  {[222]} &(2, 2) & 0,2,3,4,6         \\
\cline{4-5}    &              &       &(2, 0)&2,4               \\
\cline{4-5}    &              &       &(0, 0)&0                 \\
\cline{2-5}    & \raisebox{-7.50ex}[0cm][0cm]
{[2]$\otimes$[1111]}&
\raisebox{-4.00ex}[0cm][0cm]  {[3111]} & (3, 0) & 0,3,4,6         \\
\cline{4-5}    &              &        & (2,1)  & 1,2,3,4,5   \\
\cline{4-5}    &              &        & (1, 0) & 2                \\
\cline{3-5}    &              &
\raisebox{-3.0ex}[0cm][0cm]  {[2]}    & (2, 0) & 2,4              \\
\cline{4-5}    &              &        & (0, 0) & 0               \\
\hline
\end{tabular}
\end{footnotesize}
\end{center}
\end{table}

\begin{table}[htbp]
\begin{flushleft}
\textbf{\small{(Table 2 continued) Part B. negative parity states}
}
\end{flushleft}
\begin{center}
\begin{small}
\begin{tabular}
{|c|c|c|c|p{146pt}|} \hline $U(15)\otimes U(10)$  & $SU_{sdg}(5)
\otimes SU_{pf}(5)$ & $SU_{spdfg}(5)$  & O(5)  & \hspace*{20mm}
O(3) \\  \hline $[n_{sdg}] \otimes [n_{pf}]$ & $[n]^+ \otimes
[n]^-$ & $[n]$
  & $(\nu_1, \nu_2)$ & \hspace*{22mm} $L$  \\
\hline \raisebox{-37.00ex}[0cm][0cm]{[2]$ \otimes $[1]}&
\raisebox{-17.00ex}[0cm][0cm]{[4]$ \otimes $[11]}&
\raisebox{-8.50ex}[0cm][0cm]{[51]}&
                 (5, 1) & 1,2,$3^2,4^2,5^3,6^2,7^3,8^2,9^2$,10,11  \\
\cline{4-5} & & &(4, 0) & 2,4,5,6,8   \\
\cline{4-5} & & &(3, 1) & 1,$2,3^2,4,5^2$,6,7   \\
\cline{4-5} & & &(2, 0) & 2,4        \\
\cline{4-5} & & &(1, 1) & 1,3         \\
\cline{3-5} & & \raisebox{-6.50ex}[0cm][0cm]{[411]}&
                 (4, 1)& 1,2,$3^2,4^2,5^2,6^2,7^2$,8,9 \\
\cline{4-5} & & &(3, 1) & 1,$2,3^2,4,5^2$,6,7  \\
\cline{4-5} & & &(2, 1) & 1,2,3,4,5  \\
\cline{4-5} & & &(1, 1) & 1,3        \\
\cline{2-5} & \raisebox{-20.00ex}[0cm][0cm]{[22]$\otimes$[11]}&
\raisebox{-4.50ex}[0cm][0cm]{[33]}
                &(3, 3) & 1,$3^2$,4,5,6,7,9   \\
\cline{4-5} & & &(3, 1) & 1,$2,3^2,4,5^2$,6,7   \\
\cline{4-5} & & &(1, 1) & 1,3              \\
\cline{3-5} & & \raisebox{-10.50ex}[0cm][0cm]{[321]}& (3, 2) & 1,$2^2,3,4^2,5^2,6,7$,8   \\
\cline{4-5} & & &(3, 1) & 1,$2,3^2,4,5^2$,6,7  \\
\cline{4-5} & & &(2, 2) & 0,2,3,4,6  \\
\cline{4-5} & & &(2, 1) & 1,2,3,4,5  \\
\cline{4-5} & & &(2, 0) & 2,4        \\
\cline{4-5} & & &(1, 1) & 1,3        \\
\cline{3-5} & & \raisebox{-2.00ex}[0cm][0cm]{[2211]}&
                 (2, 1) & 1,2,3,4,5  \\
\cline{4-5} & & &(1, 1) & 1,3        \\
\hline \raisebox{-8.00ex}[0cm][0cm]{[0]$ \otimes $[3]}&
\raisebox{-4.00ex}[0cm][0cm]{[0]$ \otimes $[33]}&
\raisebox{-4.00ex}[0cm][0cm]{[33]}&
                 (3, 3) & 1,$3^2$,4,5,6,7,9 \\
\cline{4-5} & & &(3, 1) & 1,$2,3^2,4,5^2$,6,7  \\
\cline{4-5} & & &(1, 1) & 1,3           \\
\cline{2-5} & \raisebox{-2.00ex}[0cm][0cm]{[0]$ \otimes $[2211]}&
\raisebox{-2.00ex}[0cm][0cm]{[2211]}&
                 (2, 1) & 1,2,3,4,5    \\
\cline{4-5} & & &(1, 1) & 1,3           \\
\hline
\end{tabular}
\end{small}
\end{center}
\end{table}

With a chosen set of parameters $\epsilon$=0.3~MeV,
${\epsilon}^{\prime}=0.7$~MeV, $A={A}^{\prime}=0$, $B=-0.06$~MeV,
$C=0.05$~MeV, $D=0.006$~MeV, we get the energy spectra with
positive, negative parity as shown in Fig.~1, Fig.~2,
respectively.

Looking through the figures, one can easily know that much more
energy bands than those in the $sd$ IBM and the $spdf$ IBM appear
here and the band structure is more complicated than that in the
previous models. Only for the subset with positive parity, there
exist vibrational bands with angular momentum sequence $\Delta I
=4$ belonging to the totally symmetric irrep $[6]$ of the SU(5)
group. Meanwhile, the non-totally symmetric irrep $[4,2]$
generates a rotational bands with angular momentum sequence
$\Delta I =2$. In this band, the states with angular momentum $I=4
k$ ($k$ is an integer) are generated by the totally symmetric
irreps of the SO(5) group, while those with $I=4 k +2 $ are
generated by the non-totally symmetric irreps of the SO(5). With
the definition of the transition energy $E_{\gamma}(I) = E(I) -
E(I\!-\!2)$, one can readily get the contribution of the term with
the SO(5) symmetry to the $\gamma$-ray energy $E_{\gamma}$ as
$E_{\gamma}^{SO(5)} ([I]=4k \! + \! 2) = 6 C$ and
$E_{\gamma}^{SO(5)}([I] = 4k) = (4 [\frac{I}{2}] \! - \! 4) C $.
Therefore the term with the SO(5) symmetry makes the $E_{\gamma}$
staggering even though the $\vert C \vert$ may be very small.
There involves then $\Delta I =2$ energy staggering in such a
band. The rotational bands with $\Delta I =4$ bifurcation in both
ND states and SD states may thus be attributed to that the nuclei
may possess the $\mbox{SU}_{spdfg}(5)$ symmetry. On the other
hand, combining the spectrum with positive parity and that with
negative parity, one may know that there display bands with
angular momentum sequence $2^{+}$, $3^{-}$, $4^{+}$, $5^{-}$,
$\cdots$, or $2^{-}$, $3^{+}$, $4^{-}$, $5^{+}$, $\cdots$, which
is usually referred to as ``simplex symmetry". The appearance of
such kind bands is just the feature of the nucleus with both
octupole deformation and quadrupole, hexadecupole deformations.

\section{Summary and Remarks}

\hspace*{6mm}In this paper, the SU(5) dynamical symmetry of the $spdfg$ IBM has been discussed. The algebraic
structure such as the generators, Casimir operators, branching rules of the irreducible representation reductions,
and the typical energy spectrum are obtained. It shows that many bands that do not exist in the $sd$ IBM and
$spdf$ IBM appear in the $\mbox{SU}_{spdfg}(5)$ symmetry and the band structure is more complicated. Of particular
interest are the existence of both vibrational bands generated by the totally symmetric irrep of the
$\mbox{SU}_{spdfg}(5)$ group and rotational bands generated by the non-totally symmetric irrep. The bands with
$\Delta I =4$ bifurcation in the normally deformed states and superdeformed states may be attributed to that the
nuclei may hold the $SU_{spdfg}(5)$ symmetry. Moreover, the simultaneous appearance of the vibrational bands and
rotational bands indicates that the $SU(5)$ symmetry in $spdfg$ IBM may be applicable to describe the shape
coexistence and phase transition in nuclei.

On the other hand, it is known that electromagnetic transition
rates are important signatures of nuclear structure. In general,
with only the one-body operators being taken into account, the E2,
E1, E3, M1 transition (which are believed to be the more important
ones) operator can be given as
\begin{eqnarray}
\hat{T}(E2)&=&e_{2}\{[s^{\dag}\widetilde{d}+d^{\dag}\widetilde{s}]^{(2)}
+\chi_{dd}^{2}[d^{\dag}\widetilde{d}]^{(2)} + \chi_{gg}^{2}[g^{\dag}
\widetilde{g}]^{(2)} + \chi_{dg}^{2}[d^{\dag}\widetilde{g} +
g^{\dag}\widetilde{d}]^{(2)}\nonumber \\
& & + \chi_{pp}^{2}[p^{\dag}\widetilde{p}]^{(2)}+
\chi_{ff}^{2}[f^{\dag}\widetilde{f}]^{(2)} +
\chi_{pf}^{2}[p^{\dag}\widetilde{f}+f^{\dag}\widetilde{p}]^{(2)}
\, \},
\end{eqnarray}
\begin{equation}
\hat{T}(E1)=e_{1}\{[s^{\dag}\widetilde{p} \! + \!
p^{\dag}\widetilde{s}]^{(1)} \! + \!
\chi_{pd}^{1}[p^{\dag}\widetilde{d} \! + \! d^{\dag} \widetilde{p}
]^{(1)} \! + \! \chi_{df}^{1}[d^{\dag} \widetilde{f} \! + \!
f^{\dag} \widetilde{d}]^{(1)} \! + \!
\chi_{fg}^{1}[f^{\dag}\widetilde{g}\! + \!
g^{\dag}\widetilde{f}]^{(1)} \, \},
\end{equation}
\begin{equation}
\hat{T}(E3)=e_{3}\{[s^{\dag}\widetilde{f} \! + \!
f^{\dag}\widetilde{s}]^{(3)} \! + \!
\chi_{pd}^{3}[p^{\dag}\widetilde{d} \! + \! d^{\dag} \widetilde{p}
]^{(3)} \! + \! \chi_{df}^{3}[d^{\dag} \widetilde{f} \! + \!
f^{\dag} \widetilde{d}]^{(3)} \! + \!
\chi_{fg}^{3}[f^{\dag}\widetilde{g}\! + \!
g^{\dag}\widetilde{f}]^{(3)} \, \},
\end{equation}
\begin{equation}
\hat{T}(M1)=g_{L} \hat{L} \, ,
\end{equation}
where $\chi_{\alpha\beta}^{k}(\alpha,\beta=p,d,f,g)$ are the
structure constants, $e_{k} (k = 1,2,3)$ are the effective charge,
$\hat{L}$ is the angular momentum operator as shown in Eq.~(21),
$g_{L}$ is the rotational g-factor . After analysis, the selection
rules can be expressed as $\Delta n_{sdg} = 0$ with $\Delta
n_{d}=\pm 1,0$ and/or $\Delta n_{g}=\mp 1,0$ or $\Delta n_{pf}=0$
with $\Delta n_{p}=0, \pm1$ and/or $\Delta n_{f}=0,\mp1$ for the
E2 transition. The selection rules for the E1 and E3 transitions
can be given as $\Delta n_{sdg} = \pm 1$ with $\Delta n_{s}=\pm 1$
and/or $\Delta n_{d}=\pm 1$ and/or $\Delta n_{g} = \pm 1$ and
$\Delta n_{pf}=\mp 1$ with $\Delta n_{p}= \mp 1$ and/or $\Delta
n_{f}=\mp 1$. However the calculation of the reduced transition
rates are so complicated that we postpone it now.

\bigskip

This work is supported by the National Natural Science Foundation
of China under the contract No. 19875001, the Foundation for
University Key Teacher by the Ministry of Education, China and the
Founds of the Key Laboratory of Heavy Ion Physics at Peking
University, Ministry of Education, China. Helpful discussions with
Professors H.Z. Sun and Q.Z. Han are acknowledged with thanks.

\begin{center}
\begin{figure}
\includegraphics[scale=0.6,angle=0]{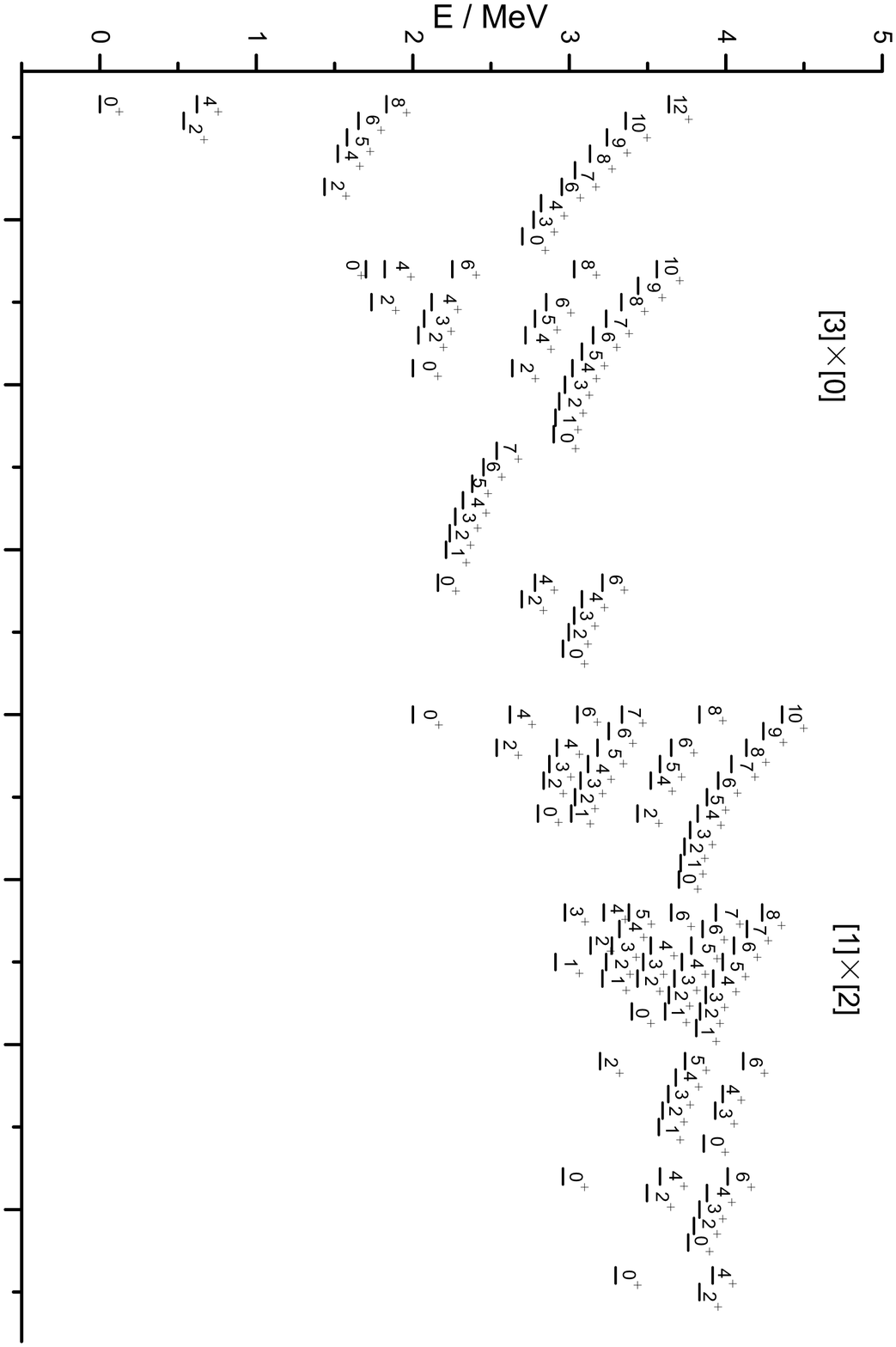}
\caption{Energy spectrum of the positive parity states
with total boson number $N=3$} 
\end{figure}
\end{center}

\begin{center}
\begin{figure}
\includegraphics[scale=0.6,angle=0]{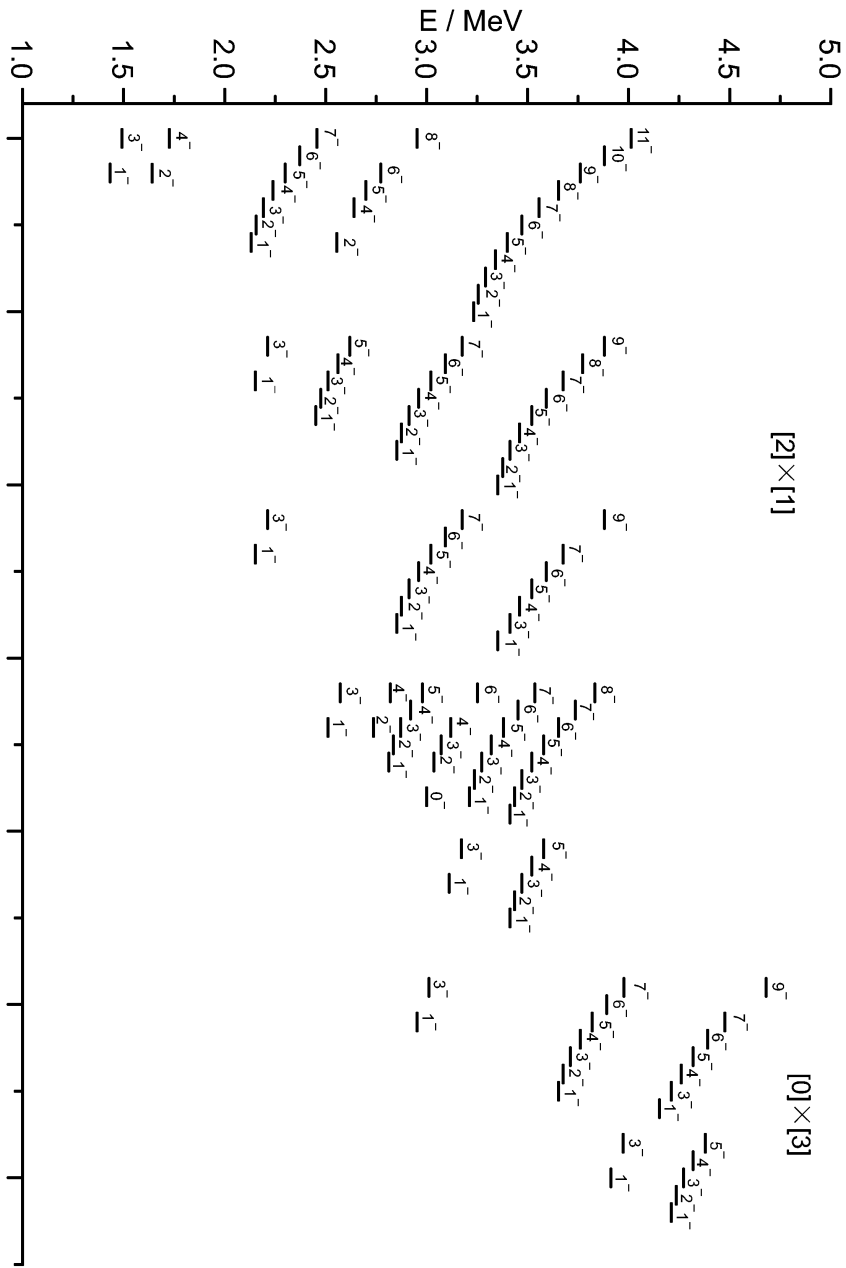}
\caption{Energy spectrum of the negative parity states with total
boson number $N=3$}
\end{figure}
\end{center}

\end{document}